# Computing the Expected Value of Sample Information Efficiently: Expertise and Skills Required for Four Model-Based Methods

Natalia R. Kunst, MSc,[1,2,3,4] Edward CF Wilson, PhD,[5] Fernando Alarid-Escudero, PhD,[6] Gianluca Baio, PhD,[7] Alan Brennan, PhD,[8] Michael Fairley, BE(Hons),[9] David Glynn, PhD,[10] Jeremy D. Goldhaber-Fiebert, PhD,[9] Chris Jackson, PhD,[11] Hawre Jalal, MD, PhD,[12] Nicolas A. Menzies, PhD,[13] Mark Strong, PhD,[8] Howard Thom, PhD,[14] Anna Heath, PhD,[15,16,7] and on behalf of the Collaborative Network for Value of Information (ConVOI)

[1]University of Oslo; [2]Yale University School of Medicine; [3]Amsterdam UMC; [4]LINK Medical Research; [5]Health Economics Group, Norwich Medical School, University of East Anglia; [6]Center for Research and Teaching in Economics (CIDE); [7]University College London; [8]School of Health and Related Research (ScHARR), University of Sheffield; [9]Stanford Health Policy, Centers for Health Policy and Primary Care and Outcomes Research, Stanford University; [10]University of York; [11]MRC Biostatistics Unit, University of Cambridge; [12]University of Pittsburgh; [13]Harvard TH Chan School of Public Health; [14]University of Bristol; [15]The Hospital for Sick Children; [16]University of Toronto

## Abstract

### Objectives

Value of information (VOI) analyses can help policy-makers make informed decisions about whether to conduct and how to design future studies. Historically, a computationally expensive method to compute the Expected Value of Sample Information (EVSI) restricted the use of VOI to simple decision models and study designs. Recently, four EVSI approximation methods have made such analyses more feasible and accessible. We provide practical recommendations for analysts computing EVSI by evaluating these novel methods.

### Methods

Members of the Collaborative Network for Value of Information (ConVOI) compared the inputs, analyst's expertise and skills, and software required for four recently developed approximation methods. Information was also collected on the strengths and limitations of each approximation method.

### Results

All four EVSI methods require a decision-analytic model's probabilistic sensitivity analysis (PSA) output. One of the methods also requires the model to be re-run to obtain new PSA outputs for each EVSI estimation. To compute EVSI, analysts must be familiar with at least one of the following skills: advanced regression modeling, likelihood specification, and Bayesian modeling. All methods have different strengths and limitations, e.g., some methods handle evaluation of study designs with more outcomes more efficiently while others quantify uncertainty in EVSI estimates. All methods are programmed in the statistical language R and two of the methods provide online applications.

### Conclusion

Our paper helps to inform the choice between four efficient EVSI estimation methods, enabling analysts to assess the methods' strengths and limitations and select the most appropriate EVSI method given their situation and skills.

## Highlights

- The Expected Value of Sample Information (EVSI) can be used to prioritize research and design future studies to reduce decision uncertainty for policy-makers. Four recently published methods have overcome the computational issues associated with EVSI analysis but practical guidance on using and distinguishing these methods is lacking.
- Because the four methods use different approaches to estimate EVSI, requiring different expertise and skills, members of the Collaborative Network on Value of Information (ConVOI) reviewed these four EVSI computation methods to understand their required skills and inputs. They also identified the strengths and limitations of these methods and provide step-by-step guides.
- By comparing these methods, ConVOI provides practical guidance for analysts looking to compute EVSI and use it for study design. Analysts now have useful information to confidently select the most appropriate EVSI estimation method for their application and expertise.

**Introduction**

Decisions on which research studies to fund are intrinsically economic in nature; that is, public, private and third sector (research charity) funders have finite resources and their decisions bear an opportunity cost. Therefore, it is important to prioritise research studies that are expected to yield the greatest benefit for every dollar or euro spent.

Typically, research priorities are set through consultation with experts, decision- or policy-makers and other key stakeholders (e.g., the James Lind Alliance [1]) [2]. Study outcomes are then selected for a number of reasons, such as ensuring timely collection of data. Most research designs focus on clinical outcomes and rarely consider economic implications. The sample size for most randomized controlled trials (RCTs) is determined using power calculations which explicitly manage type I and II errors for a statistical test of the selected primary outcome. However, this approach fails to take into account the opportunity cost of research: a large (and expensive) trial could be proposed while that knowledge could have minimal value to society. In contrast, Value of Information analysis (VOI) values an RCT and other types of research studies in terms of how much they reduce decision uncertainty (i.e., the probability of making a sub-optimal decision) about the best treatment for use in a population of interest. If a sub-optimal decision is made, then potential health improvements are foregone. The probability of making the wrong decision is multiplied by the size of the loss incurred by that decision to generate the expected loss associated with making a sub-optimal decision [3]. New data are expected to reduce decision uncertainty, and so reduce the expected loss. The Expected Value of Sample Information (EVSI) measures the "expected reduction in expected loss" from a given research study. Scaled up to the relevant population this can be expressed in health terms of life years, or quality-adjusted life years (QALYs), or in monetary units [4, 5]. The difference between the EVSI and the cost of the research is the Expected Net Benefit of Sampling (ENBS).

An increasing number of authorities and Health Technology Assessment (HTA) agencies acknowledge the importance of VOI as part of a cost-effectiveness analysis and are recommending it [6-9]. Thus, EVSI methods will soon be required from analysts who conduct cost-effectiveness analyses for HTA agencies. Until recently, EVSI calculations were extremely computationally expensive, potentially taking weeks or months [10], as they required nested simulation methods [11]. Hence, most VOI analyses were restricted to computing the Expected Value of Perfect Information (EVPI) or the Expected Value of Partial Perfect Information (EVPPI). However, new algorithms and associated software [12-20] enable the efficient computation of EVSI for realistic decision models.

While these methods have lowered the computational barriers to VOI analyses, they differ in their approach, requiring different expertise and skills. However, there is no structured comparison of the practical steps required to use them. Thus, it is challenging for analysts to determine which method is appropriate for their situation and expertise. Additionally, each method has different strengths and limitations that could make it more suitable for a given decision problem. The Collaborative Network for Value of Information (ConVOI) [21] is an international group of researchers with interests in the application and development of methods for VOI calculation. This manuscript provides practical guidance and good practice recommendations for computing EVSI using four recently developed approximation methods. For each method, we: 1) provide a step-by-step guide to its use, 2) compare the expertise and skills required to implement it, and 3) highlight its strengths and limitations.

**Four EVSI approximation methods**

This paper focuses on four recently developed estimation methods, developed by Strong, Oakley (13) (regression-based method [RB]), Menzies (14) (importance sampling method [IS]), Jalal and Alarid-Escudero (16) building on work from Jalal, Goldhaber-Fiebert (15) (Gaussian approximation method [GA]), and Heath, Manolopoulou (17) (moment matching method [MM]). These calculation methods were selected as they place limited restrictions on the complexity of the underlying decision-analytic model and/or data collection exercise when calculating EVSI. The recommendations presented in this paper were developed and reviewed by ConVOI to aid analysts looking to compute EVSI.

**Inputs Required for Efficient EVSI Calculation**

The considered EVSI approximation methods have diverse requirements and make different assumptions. However, all require a decision-analytic model on which a probabilistic sensitivity analysis is conducted. The IS and MM methods also require calculation of the EVPPI for the parameter(s) to be evaluated and updated in the proposed study. If studies informing different parameters or groups of parameters are considered, then EVPPI would need to be computed for each. The following section briefly outlines these requirements and assumptions.

*Decision-Analytic Model*

VOI requires an objective function that should be optimized to determine the best course of action [22]. Economic evaluations in health care typically define this in terms of net health or monetary benefit for each of $T$ interventions [23], which uses a "willingness-to-pay" threshold to put the health consequences on the same scale as the costs for the different interventions [23]. The function that computes this usually takes the form of a decision-analytic model.

Decision-analytic models are mathematical models often used for cost-effectiveness analyses. They draw on set of inputs that we denote $\boldsymbol{\theta}$. These inputs could include information on the prevalence of a disease, the effectiveness of treatments, background mortality, health-related utility weights and costs. The model can take a number of forms, foremost among them are decision trees, Markov models and microsimulation models [24]. Conceptually, the decision-analytic model maps a set of inputs to the output, net benefit (either in monetary or health units). Assuming risk neutrality, the intervention with the highest expected net benefit should be implemented in the wider population.

*Probabilistic Sensitivity Analysis*

Probabilistic sensitivity analysis (PSA), sometimes known as probabilistic analysis or uncertainty analysis, is performed to propagate input parameter uncertainty to the model output under each decision alternative and thereby quantify decision uncertainty in a cost-effectiveness analysis [25]. PSA simultaneously varies all parameters for which there is meaningful uncertainty. Uncertainty in the model inputs is characterized using probability distributions, $p(\boldsymbol{\theta})$ [26].

PSA is often conducted using Monte Carlo methods where $S$ parameter sets are drawn from $p(\boldsymbol{\theta})$, for the whole set of model parameters $\boldsymbol{\theta} = (\theta_1, \ldots, \theta_P)$. The decision-analytic model is evaluated at all $\boldsymbol{\theta}_s$, $s = 1, \ldots, S$ to estimate the costs and health outcomes of each of the $T$ strategies. This produces a distribution for the net monetary benefit for each strategy, which we denote $\text{NB}_t^{\boldsymbol{\theta}}$, for $t = 1, \ldots, T$. In non-linear models, PSA is required to generate the expected net monetary benefit of each treatment strategy, $\text{E}_\theta[\text{NB}_t^{\boldsymbol{\theta}}]$. PSA results can be presented in the form of a cost-effectiveness acceptability curve (CEAC), cost-effectiveness acceptability frontier (CEAF), expected loss curves (ELCs), and cost-effectiveness plane [3, 6]. Furthermore, PSA is key to determining the value of potential future research using VOI methods.

To compute EVSI, we require the PSA simulations of the model inputs $\boldsymbol{\theta}_s$ and the corresponding simulations for the net monetary benefit $\text{NB}_t^{\boldsymbol{\theta}_s}$ for $s = 1, \ldots, S$ and $t = 1, \ldots, T$. These simulations should be saved in a matrix or spreadsheet form, often called a PSA dataset,[1] where the columns contain first the input parameters and then the outputs from a decision-analytic model. Each row then contains the parameter sets drawn from their distributions, and their corresponding simulated results from a decision-analytic model output. We have provided an example of the PSA simulations in Supplementary Materials.

While PSA results are necessary for all the four efficient EVSI calculation methods, two of the methods (RB and GA) require the above mentioned PSA matrix from a traditional cost-effectiveness analysis while the other two (IS and MM) require an augmented PSA simulation matrix that is presented in the following section.

*Expected Value of Partial Perfect Information*

The EVPPI computes the value of eliminating *all* uncertainty about a subset of the model parameters [27-29]. Specifically, the model parameters are split into two subsets $\boldsymbol{\theta} = (\boldsymbol{\phi}, \boldsymbol{\psi})$, where we propose to gather further information about the model parameters $\boldsymbol{\phi}$. Typically, a proposed study does not collect information about all the underlying parameters in a decision-analytic model and therefore, $\boldsymbol{\psi}$ indicates the parameters that will not be directly informed by the proposed data collection. Mathematically, the EVPPI for the parameters $\boldsymbol{\phi}$ is defined as

$$\text{EVPPI} = \text{E}_\phi\left[\max_t \text{E}_{\psi|\phi}\left[\text{NB}_t^{\boldsymbol{\theta}}\right]\right] - \max_t \text{E}_\theta\left[\text{NB}_t^{\boldsymbol{\theta}}\right], \qquad (1)$$

where the inner expectation in the first term of equation (1) calculates the net monetary benefit for each intervention conditional on $\boldsymbol{\phi}$. The second term calculates the value of the decision made with current information, i.e., the expected net benefit of the treatment with the highest expected net benefit.

---

[1] The word "dataset" is not used in the traditional sense; here, the PSA dataset simply contains simulated values from distributions representing the uncertainty in the parameter estimates.

The PSA outcomes are simulated using probability distributions for $\phi$ and the EVPPI is estimated by computing the net monetary benefit for each intervention conditional $\phi = \phi_s$ for $s = 1, \ldots, S$. Several approximation methods are available to estimate the conditional net monetary benefit for each value $\phi_s$ [27, 30-34] and have been formally assessed in [28] and [29]. Software is also available to perform these calculations [35-37]. The MM and IS methods require a PSA simulation matrix with $T$ additional columns that contain Monte Carlo estimates of the expected net benefits $\mathrm{E}_{\psi|\phi}[\mathrm{NB}_t^{\theta}]$, conditional on $\phi_s$ for $s = 1, \ldots, S$, but averaged over the uncertain values of $\psi$. We denote these simulations $\eta_t^s$.

**EVSI Methods**

EVSI for a proposed research strategy that collects additional data, denoted $X$, is defined as

$$\mathrm{EVSI} = E_X\left[\max_t \mathrm{E}_{\theta|X}[\mathrm{NB}_t^{\theta}]\right] - \max_t \mathrm{E}_{\theta}[\mathrm{NB}_t^{\theta}]. \qquad (2)$$

In this setup, $X$ are observ*able*, but not yet observ*ed* — and possibly never will be. At this stage, we consider the possibility of collecting data in the future and we compute EVSI to determine whether we should defer making a decision on the optimal intervention, among the $T$ possible options, and instead invest money and time collecting $X$.

In line with a full Bayesian approach (which underpins the ideas behind the VoI analysis), the distribution of $X$ is defined by $p(X, \theta) = p(\theta)p(X|\theta)$ where $p(\theta)$ is the marginal distribution of $\theta$ and $p(X|\theta)$ is the sampling distribution of the data. In a full Bayesian setting, $p(\theta)$ is a "prior" distribution — in the sense that it represents the current level of uncertainty on the model parameters, *before* observing the new data $X$. In reality, this distribution is defined by the PSA procedure and can represent the result of a Bayesian update given observed data that is used to construct the current economic model.

We note that $X$ will give information on the subset of model parameters $\phi$, where $\phi$ could be the whole set $\theta$. By definition, this means that $\psi$ and $X$ are independent given $\phi$ and that $p(X|\theta) = p(X|\phi)$.

The second term of equation (2) can be estimated from the initial PSA. To obtain simulations for $X$, we simulate potential study outcomes, $X_s$, $s = 1, \ldots, S$, for each row of the PSA dataset, i.e., we generate a single sample from $p(X|\theta^s)$ for $s = 1, \ldots, S$. Traditionally, EVSI has been computationally demanding and methodologically challenging because a large number of simulations are needed to estimate the posterior mean of the net benefit for each $X_s$. This requires a Bayesian model for the distribution of the data and the parameters, where the posterior mean is given by

$$\mu_t(X_s) = E_{\theta|X^s}[NB_t^\theta]. \qquad (3)$$

The EVSI approximation methods in this paper estimate $\mu_t(X)$ with a reduced computational burden. For computational simplicity and stability, VOI calculations should be undertaken using *incremental* net benefit defined, without loss of generality, as

$$INB_t^\theta = NB_{t*}^\theta - NB_t^\theta,$$

for $t = 2, \ldots, T$. We could also use the opportunity loss, defined as the *incremental* net benefit of all treatments from the optimal strategy $t^*$. By estimating the posterior mean of the incremental net benefit, we only need to add $T - 1$ columns containing the posterior mean conditional on $X_s$ to our PSA dataset. Once these simulations, denoted $\mu_t(X_s)$, are available, the EVSI can be estimated by

$$\frac{1}{S}\sum_{s=1}^{S} \max_t\{\mu_t(X_s)\} - \max_t\left\{\frac{1}{S}\sum_{s=1}^{S} \mu_t(X_s)\right\}, \qquad (4)$$

where $\mu_1(X_s) = 0$ for all $s = 1, \ldots, S$.

**Expertise Required for EVSI Calculation**

We have identified three skills that the analyst may require to compute EVSI using the approximation methods reviewed in this manuscript. None of the methods require all three skills but each method requires the mastery of at least one of the skills presented. Table 1 summarises the required skills and inputs for each method.

*Regression Methods*

Regression methods that model the relationship between a set of predictors and an outcome of interest are required for both the RB and GA methods. Both methods use regression metamodeling to model the incremental net benefit as a function of model inputs or quantities related to model inputs. The two methods differ as they require alternative covariates.

Both of these EVSI methods model relatively complex relationships using models that require little information about the functional relationship between the independent and dependent variables. To account for potential nonlinear relationships between these variables, flexible regression methods characterized using Generalized Additive Models (GAMs) [38] are the predominant regression methods for EVSI calculation. Standard software is available to fit GAMs [39-41], but the EVSI estimate can be affected by the structure of the GAM model and, thus, an understanding of these models is required. Gaussian Processes [42] have also been suggested for more complex problems [13], so knowledge of these methods may also be required.

If regression methods are used to estimate EVSI, it is important to assess whether the regression model has correctly captured the relationship between the independent and dependent variables [13]. Regression methods assume that the residual error is uncorrelated with the fitted values from the regression, has zero mean and trivial covariance structure. These assumptions can be confirmed by observing no groups, systematic features, or outliers in the plot of the residuals against the fitted values [43].

*Specification of data generating distribution*

EVSI computes the expected value of collecting information in a future research study. The future research study would collect a potential data set, $X$, with $N_O$ different clinical, health or economic outcomes from $N$ participants. For example, a clinical trial would collect the primary and secondary outcomes from each participant.

All EVSI computation methods require the simulation of potential data sets (the GA method requires study data simulation to compute the effective sample sizes of the priors of the $\phi$ only if these cannot be estimated directly, c.f. Bayesian Updating). As outlined above, this is achieved by specifying the assumed data generating distribution $p(X|\phi)$. Thus, analysts must have knowledge of probability distributions.

For the RB method, the simulated data sets $X_s$ for $s = 1, \dots, S$ must be summarized to reflect how the data would be analyzed following the trial. This requires expertise in standard methods for the statistical analysis of study results, for example, maximum likelihood procedures. A variation of the GA method can be also be applied if a summary statistic is available.

For the IS method, the analytic likelihood for $X$ must be specified and coded as a function. Likelihood is used in its proper statistical meaning and, therefore, this specification requires statistical expertise. Typically, this likelihood specification will require an analyst who has knowledge of statistical theory, function development and coding.

*Bayesian Updating*

Bayesian methods are a statistical paradigm in which conclusions are formally updated as evidence accumulates [44]. A full Bayesian model consists of probability distributions specified for the model inputs $p(\boldsymbol{\theta})$, the prior distribution defined using the PSA distributions, and the sampling distribution for the proposed data collection. Based on these distributions, Bayesian methods update the distribution of the parameters conditional on the data $\boldsymbol{X}$ to produce a posterior distribution. In practice, this is undertaken using specialized software such as BUGS [45], JAGS [46] or Stan [47]. The MM and, in some cases, GA methods are explicitly based on Bayesian analysis, which requires expertise in these programs and in the specification of Bayesian models. This is advantageous if the final trial analysis will be performed according to Bayesian principles as the statistical analysis plan has been developed as part of the design process.

Additionally, the GA method is based on the prior effective sample size [48]. The prior effective sample size, denoted $n_0$, is the number of participants that would need to be studied to obtain the level of information in the prior. In some Bayesian models, typically based on conjugate prior-likelihood pairs, the prior effective sample size can be estimated directly from the parameters of the distributions [48]. If this is not possible, then Jalal and Alarid-Escudero (16) present two algorithms to compute the prior effective sample size (see the Supplementary Materials). The first of these is based on summary statistics and the second on a full Bayesian analysis.

Table 1. The skills and inputs required to compute EVSI with the regression-based, importance sampling, Gaussian approximation and moment matching methods

| # | Requirements | RB | IS | GA | MM |
|---|---|---|---|---|---|
| | *Inputs* | | | | |
| 1 | Decision-analytic model | | | | ✓ |
| 2 | Probabilistic sensitivity analysis | ✓ | ✓ | ✓ | ✓ |
| 3 | Simulations of the expected net benefit conditional on $\phi$ (required to compute EVPPI) | | ✓ | | ✓ |

| | *Skills* | | | | |
|---|---|---|---|---|---|
| 1 | Regression methods | | ✓ | | ✓ |
| 2 | Specification of likelihoods | | ✓ | ✓ | * |
| 3 | Bayesian updating | | | * | ✓ |

GA, Gaussian approximation method; IS, importance sampling method; MM, moment matching method; RB, regression-based method; ✓ indicates that the skill/input is required; * indicates that the skill/input *may be* required.

**Calculating EVSI**

To highlight how these skills and inputs are used to compute EVSI, we have presented step-by-step guides to each of the four methods considered. These algorithms are provided in Boxes 1-4.

Box 1. Step-by-step guide for regression-based method

> **Box 1: Regression-Based Method**
> 1. Perform PSA simulation to obtain $\boldsymbol{\theta}_s$ and $\text{INB}_t^{\boldsymbol{\theta}_s}$, $s = 1, \ldots, S$.
> 2. For each $s = 1, \ldots, S$:
>    (a) Simulate a dataset $\boldsymbol{X}_s$ from $p(\boldsymbol{X} \mid \boldsymbol{\phi}_s)$.
>    (b) Summarise this dataset, producing the quantity or quantities that would be estimated from it in a trial, such as the mean. We denote this summary of the dataset $W(\boldsymbol{X}_s)$.
> 3. Fit $T - 1$ regression models with $\text{INB}_t^{\boldsymbol{\theta}_s}$ as the outcome and $W(\boldsymbol{X}_s)$ as the covariates.
> 4. Extract the fitted values from these regressions to estimate $\mu_t(\boldsymbol{X}_s)$.
>
> The downloadable R package and R codes for the regression-based method are available at http://savi.shef.ac.uk/SAVI/.

INB, incremental net benefit; PSA, probabilitic sensitivity analysis; θ, set of inputs in a decision-analytic model; p(θ), probability distributions to characterize uncertainty in the model inputs; S, number of parameter sets that are drawn from p(θ) in PSA; T, number of considered interventions; θ=(φ,ψ), φ are model parameters for which we are aiming to collect further information and ψ are parameters that will not be directly informed by the proposed data collection; X, new data proposed to be collected; μ, posterior mean; W(X), a summary measure for the data.

Box 2. Step-by-step guide for importance sampling method

---

**Box 2: Importance Sampling Method**

1. Perform PSA simulation to obtain $\boldsymbol{\theta}_s$ and $\text{INB}_t^{\boldsymbol{\theta}_s}$, $s = 1, \ldots, S$.

2. Estimate EVPPI to obtain simulations of $\eta_t^s$, the inner expectation in the first term of equation:

$$\text{EVPPI} = \text{E}_{\boldsymbol{\phi}} \left[ \max_t \text{E}_{\boldsymbol{\psi}|\boldsymbol{\phi}} \left[ \text{NB}_t^{\boldsymbol{\theta}} \right] \right] - \max_t \text{E}_{\boldsymbol{\theta}} \left[ \text{NB}_t^{\boldsymbol{\theta}} \right].$$

3. For each $s = 1, \ldots, S$:

   (a) Simulate a dataset $\boldsymbol{X}_s$ from $p(\boldsymbol{X} \mid \boldsymbol{\phi}_s)$.
   
   (b) Compute the likelihood of $\boldsymbol{X}_s$ conditional on all PSA simulations for $\boldsymbol{\theta}$, $L_r$, $r = 1, \ldots, S$.
   
   (c) Compute $l_r = \frac{L_r}{\sum_{r=1}^{S} L_r}$ so $l_r$ sums to 1.
   
   (d) Calculate the weighted sum of $\eta_t^r$, $\sum_{r=1}^{S} l_r \eta_t^r$.

4. Each weighted sum estimates $\mu_t(\boldsymbol{X}_s)$.

---

INB, incremental net benefit; EVPPI, Expected Value of Partial Perfect Information; PSA, probabilitic sensitivity analysis; θ, set of inputs in a decision-analytic model; p(θ), probability distributions to characterize uncertainty in the model inputs; S, number of parameter sets that are drawn from p(θ) in PSA; T, number of considered interventions; θ=(φ,ψ) φ are model parameters for which we are aiming to collect further information and ψ are parameters that will not be directly informed by the proposed data collection; X, new data proposed to be collected; μ, posterior mean; η, PSA simulations used to compute EVPPI.

Box 3. Step-by-step guide for Gaussian approximation method

**Box 3: Gaussian Approximation Method**

1. Perform PSA simulation to obtain $\boldsymbol{\theta}_s$ and $\text{INB}_t^{\boldsymbol{\theta}_s}$, $s = 1, \ldots, S$.

2. Fit $T - 1$ regression models with $\text{INB}_t^{\boldsymbol{\theta}_s}$ as outcomes and $\boldsymbol{\phi}_s$ as covariates.

3. For each $p$ of the $P$ elements in $\boldsymbol{\phi}$:

   (a) Determine the prior effective sample size $n_0^p$. Methods to estimate the prior effective sample size are presented in the supplementary material.

   (b) For a proposed study with $N$ participants, compute a weighted sum of $\boldsymbol{\phi}$ and $\bar{\boldsymbol{\phi}}$ (the mean of the parameters) by rescaling the simulations for the $p^{\text{th}}$ element of $\boldsymbol{\phi}$ by multiplying by $\frac{N}{N+n_0^p}$ and multiplying $\bar{\boldsymbol{\phi}}$ by $\left(1 - \frac{N}{N+n_0^p}\right)$.

4. Using the regression models from Step 2, predict the model outcomes for the rescaled $\boldsymbol{\phi}$ simulations.

5. The fitted values from Step 4 estimate $\mu_t(\boldsymbol{X}_s)$.

INB, incremental net benefit; PSA, probabilitic sensitivity analysis; θ, set of inputs in a decision-analytic model; S, number of parameter sets that are drawn from p(θ) in PSA; T, number of considered interventions; θ=(φ,ψ) φ are model parameters for which we are aiming to collect further information and ψ are parameters that will not be directly informed by the proposed data collection; X, new data proposed to be collected; μ, posterior mean.

Box 4. Step-by-step guide for moment matching method

> **Box 4: Moment Matching Method**
>
> 1. Perform PSA simulation to obtain $\boldsymbol{\theta}_s$ and $\text{INB}_t^{\boldsymbol{\theta}_s}$, $s = 1, \ldots, S$.
>
> 2. Estimate EVPPI to obtain simulations of $\eta_t^s$, the inner expectation in the first term of equation:
>
> $$\text{EVPPI} = \text{E}_{\boldsymbol{\phi}}\left[\max_t \text{E}_{\boldsymbol{\psi}|\boldsymbol{\phi}}\left[\text{NB}_t^{\boldsymbol{\theta}}\right]\right] - \max_t \text{E}_{\boldsymbol{\theta}}\left[\text{NB}_t^{\boldsymbol{\theta}}\right].$$
>
> 3. Extract $Q$, with $30 < Q < 50$, sample quantiles from the simulations of $\boldsymbol{\phi}$, denoted $\boldsymbol{\phi}_q$. Functions are available in the `EVSI` package in `R` to perform this step.
>
> 4. For $q = 1, \ldots, Q$:
>    (a) Simulate a future dataset from the sampling distribution $p(\boldsymbol{X} \mid \boldsymbol{\phi}_q)$.
>    (b) Use Bayesian methods to update the distribution of the model parameters.
>    (c) Rerun the probabilistic sensitivity analysis to update the distribution of the net monetary benefit.
>    (d) Calculate the *variance* of the net monetary benefit, denoted $\sigma_q^2$.
>
> 5. Calculate $\sigma^2 = \text{Var}\left[\text{NB}_t^{\boldsymbol{\theta}_s}\right] - \frac{1}{Q}\sum_{q=1}^Q \sigma_q^2$.
>
> 6. Rescale the simulations of $\eta_t^s$ so their variance is equal to $\sigma^2$.
>
> 7. These rescaled simulations estimate $\mu_t(\boldsymbol{X}_s)$.
>
> An extended version of the moment matching method estimates the EVSI for multiple alternative sample sizes for the future data for a fixed additional computational cost.

INB, incremental net benefit; EVPPI, Expected Value of Partial Perfect Information; EVSI, Expected Value of Sample Information; PSA, probabilitic sensitivity analysis; θ, set of inputs in a decision-analytic model; p(θ), probability distributions to characterize uncertainty in the model inputs; S, number of parameter sets that are drawn from p(θ) in PSA; T, number of considered interventions; θ=(φ,ψ) φ are model parameters for which we are aiming to collect further information and ψ are parameters that will not be directly informed by the proposed data collection; X, new data proposed to be collected; μ, posterior mean; η, PSA simulations used to compute EVPPI; Q, number of times PSA is performed.

**Strengths and Limitations**

As each method is different, the most suitable method for EVSI calculation will depend on the decision-analytic model, the sampling distribution for the data, the expertise of the analyst and the amount of computation time available. The following section highlights the strengths and limitations of these methods to help analysts select the most appropriate method. A summary of these strengths and limitations is presented in Table 2.

*Regression-Based Method*

*Strengths*

The decision-analytic model does not need to be rerun to produce EVSI estimates, so EVSI can be computed by an analyst with access to the PSA results only. Standard statistical software includes procedures to fit flexible regression models, particularly GAM regression, making this method relatively simple to implement once the *summary statistics* $W(X_s)$ are available. Furthermore, EVSI estimates across different sample sizes for the future data are obtained at a constant computational cost. Finally, if the model is judged to fit well, an estimate of the uncertainty in the EVSI estimate can be obtained using an algorithm developed by Strong, Oakley (13).

An online application [36] was developed to compute the EVPPI, using the regression-based method [31]. This tool, called Sheffield Accelerated Value of Information (SAVI), fits a regression model between the $\text{INB}_t^{\theta_s}$ and the parameters of interest $\phi$. Due to the similarity between the EVPPI and EVSI calculation methods, SAVI can be used to compute EVSI, once the future data sets have been summarized $W(X_s)$, $s = 1, \ldots, S$. This is achieved by augmenting the PSA matrix with (a) column(s) containing the data summaries. This must be saved and uploaded into SAVI and EVSI is then equal to the "EVPPI" calculated for the column(s) containing the data summary.

*Limitations*

For complex studies where the number of collected outcomes $N_O$ is greater than five or six, it can be challenging to fit a sufficiently accurate regression model. The simulated study data must also be correctly summarised using $W(X)$ to obtain accurate EVSI estimates. This can be challenging and time-consuming in more complex studies. The relationship between the incremental net benefit and $W(X)$ must also be well-approximated by the regression model to ensure accurate EVSI estimation.

*Importance Sampling Method*

*Strengths*

This method only requires the PSA results to compute the EVSI (i.e., access to the original decision-analytic model is not necessary). Furthermore, the IS method can be used irrespective of the number of study outcomes $N_O$ with similar complexity, once the likelihood function for the potential data to be collected has been defined.

*Limitations*

The accuracy of the EVSI estimation relies heavily on the appropriate specification of the analytic likelihood and the accurate estimation of the EVPPI. Because the most efficient EVPPI estimation methods are based on regression methods [29, 31], the analyst must be confident assessing the accuracy of regression models before proceeding with EVSI estimation. Furthermore, the IS method can have computational issues for large sample sizes of the future data as the likelihood tends to 0 with increasing sample size, leading to inaccurate EVSI estimation [10]. Finally, uncertainty in the EVSI estimation procedure cannot be estimated using this method.

**Gaussian Approximation Method**

*Strengths*

Estimation of EVSI with this method does not require to rerun the decision-analytic model; thus, access to PSA results is sufficient. This method estimates EVSI using flexible non-parametric (typically GAM) linear regression models. Although GA method requires an estimate of the prior effective sample size $n_0$ for the proposed data collection, $n_0$ can be estimated directly from previous studies. Furthermore, this method estimates EVSI across different sample sizes of the proposed data collection exercise at minimal computational cost, once the prior effective sample size has been estimated. Thus, the optimal sample size for proposed data collection can be obtained cheaply. Finally, the method can estimate the uncertainty in the EVSI estimation procedure using the same approach as in [13] based on [49].

An online repository is available for the GA method (https://zenodo.org/record/3263876), which contains a function that estimates the fitted values $\mu_t(X_s)$ from a GAM regression model [38] and estimates of $n_0^p$ for $p = 1, ..., P$. Thus, implementation for this method can be automated if estimates for the effective sample size are available for each parameter and if a GAM regression model accurately captures the relationship between the incremental net benefit and the parameters of interest.

*Limitations*

This estimation method can lead to inaccurate EVSI estimation if the prior effective sample size is small. Relatively intensive simulation methods may also be required to estimate the prior effective sample size if it cannot be obtained directly. The GA method also relies on a regression model, which may perform poorly if the regression model does not capture the relationship between the net monetary benefit and the parameters $\phi$. Moreover, if the number of collected outcomes $N_O$ is over five or six, flexible GAM regression methods become challenging [31]. Although the analyst can use linear regression model instead, this model may not be sufficiently capture the relationship, leading to inaccurate EVSI estimates.

## *Moment Matching Method*

*Strengths*

This method uses the same nested simulation structure as the gold standard nested Monte Carlo methods. This means that if an analyst has already developed the nested Monte Carlo method for EVSI estimation, it can be easily adapted for this method. Furthermore, this method has been extended [19] to estimate EVSI for multiple alternative sample sizes with a fixed additional computational cost. This extended method also provides a measure of uncertainty around the final EVSI estimate. The MM method can estimate EVSI irrespective of $N_O$ the number of outcomes considered.

The EVSI package in **R** has been developed to implement the MM method based on a Bayesian decision-analytic model. The manual is available [50] and the EVSI package can be installed in **R** using the command: devtools::install_github("annaheath/EVSI")

*Limitations*

The MM method requires simulated study data and relies on performing a PSA $Q$ times, with $30 < Q < 50$. Thus, the decision-analytic model must be rerun a significant number of times. Therefore, in computationally expensive decision-analytic models, such as microsimulation models, this method requires significant computational power. Next, if the original PSA simulation size is small, this method is inaccurate as it is based on an estimate of the variance of $\text{NB}_t^{\theta_s}$, which must be sufficiently accurate. The MM method can also lead to inaccurate EVSI estimates when the sample size of the proposed study is less than 10. Finally, this method requires an accurate estimate of EVPPI and is more accurate for studies that will have significant impact on the underlying uncertainty in the decision-analytic model, i.e., the EVPPI of $\phi$ needs to be high compared to the value of reducing all model uncertainty (i.e., EVPI), ideally greater than 40%.

Table 2. Selected strengths and limitations of the four EVSI approximation methods

| # | Characteristics | RB | IS | GA | MM |
|---|---|---|---|---|---|
| | *Strengths* | | | | |
| 1 | Estimates EVSI for complex studies collecting a large number of outcomes | | ✓ | | ✓ |
| 2 | Only requires the PSA results | ✓ | ✓ | ✓ | |
| 3 | Uses non-parametric (typically GAM) regression | ✓ | | ✓ | |
| 4 | Estimates EVSI for different study sizes with same computational cost | ✓ | | ✓ | |
| 5 | Quantifies uncertainty in estimate | ✓ | | ✓ | ✓ |
| | *Limitations* | | | | |
| 1 | Requires simulated study data | ✓ | ✓ | * | ✓ |
| 2 | Requires accurate EVPPI estimation | | ✓ | | ✓ |
| 3 | Can be computationally challenging to estimate EVSI for proposed studies with a large sample size | | ✓ | | |
| 4 | Requires simulated study data to be summarized in a low dimensional statistic | ✓ | | | |
| 5 | Struggles if proposed study has more than 5 outcomes | ✓ | | ✓ | |
| 6 | May estimate inaccurate EVSI if the proposed study has small prior effective sample size | | | ✓ | |
| 7 | May estimate inaccurate EVSI if the proposed study has small sample size | | | | ✓ |

GA, Gaussian approximation method, IS, importance sampling method, MM, moment matching method; RB, regression-based method; ✓ indicates that the characteristic is required for the given method; * indicates that the characteristic *may be* required for the given method.

## Real-World Examples

To aid the implementation of the reviewed EVSI methods, we have created a comprehensive GitHub repository that presents the code used to compute EVSI in the original publications, available at https://github.com/convoigroup. This repository also contains a suite of practical examples of EVSI computation that demonstrate EVSI calculations across several real-world examples using common decision-analytic model structures, such as Markov models. We provide both an example that was previously published in the literature [51] and hypothetical examples. These examples are all developed in the statistical computing language **R**. Thus, our GitHub repository demonstrates how the reviewed methods can be used in practice and will help analysts to implement VOI methods in their own work.

**Presenting EVSI results**

Once EVSI results are available, the EVSI package in **R** contains several graphical displays to present EVSI and related quantities to practitioners and stakeholders. Results of an EVSI analysis can be loaded into the EVSI package, irrespective of the computation method used. These graphics can then be displayed in **R** directly or explored using a dynamic graphical display launched from within **R**. For analysts unfamiliar with **R**, EVSI results can be loaded into an online interface and these graphics can be explored online at https://egon.stats.ucl.ac.uk/projects/EVSI/Test/. The optimal sample size estimated using VOI methods can also be presented in form of a curve of optimal sample size (COSS) [52].

**Conclusion**

VOI analysis has the potential to guide policy-makers in the prioritization and design of future research studies, thereby improving decision-making. Increasingly, HTA agencies are acknowledging this potential and are recommending VOI analyses to determine whether/what potential future research is needed. In this study, members of ConVOI have provided a practical guide and good practice recommendations to facilitate the implementation of these methods. Our recommendations outline the inputs, analyst skills, and software required to use each of the EVSI approximation methods. We have also highlighted the strengths and limitations of each method. These recommendations are supported by a recent review that compared properties of these four recently developed, efficient computation methods across three decision-analytic models [10].

An accompanying GitHub repository includes **R** code demonstrating all four EVSI methods in real-world examples. Thus, this guide helps analysts choose the method that is the most suitable for their application and skills, and our code repository provides practical support to aid the implementation of these methods. This increases the feasibility and accessibility of the EVSI methods as they become an important and required tool.

**Funding**

NRK was funded by the Research Council of Norway (276146) and LINK Medical Research. EW was funded by Norwich Medical School. FAE was funded by the National Cancer Institute (U01- CA-199335) as part of the Cancer Intervention and Surveillance Modeling Network (CISNET). GB was partially funded by a research grant sponsored by Mapi/ICON at University College London. AB was funded through Higher Education Funding Council for England and a range of UK and International research grant award bodies. MF was funded by the Stanford Interdisciplinary Graduate Fellowship. DG has no funding to declare. JDGF was funded in part by a grant from Stanford's Precision Health and Integrated Diagnostics Center (PHIND). CJ was funded the UK Medical Research Council programme MCUU00002/11. MS has no funding to declare. HJ was funded by NIH/NCATS grant 1KL2TR0001856. NM was supported by National Institutes of Health (NIH) [R01AI112438-02.]. HT was funded by the NIHR Biomedical Research Centre at the University Hospitals Bristol NHS Foundation Trust and the University of Bristol. AH was funded by the Canadian Institute of Health Research through the PERC SPOR iPCT grant. The funding agreement ensured the authors' independence in designing the study, interpreting the data, writing, and publishing the report.

**Contributions**

NK has made substantial contributions to: Concept and design, Acquisition of data, Analysis and interpretation of data, Drafting of the manuscript, Critical revision of paper for important intellectual content, Administrative, technical, or logistic support. EW has made substantial contributions to: Concept and design, Analysis and interpretation of data, Drafting of the manuscript, Critical revision of paper for important intellectual content. DG has made substantial contributions to: Concept and design, Acquisition of data, Analysis and interpretation of data, Critical revision of paper for important intellectual content. FAE, GB, AB, MF, JDGF, CJ, HJ, NAM, MS, and HT have made substantial contributions to: Concept and design, Analysis and interpretation of data, Critical revision of paper for important intellectual content. AH has made substantial contributions to: Concept and design, Analysis and interpretation of data, Drafting of the manuscript, Critical revision of paper for important intellectual content, Supervision. All authors approved the final draft.

# Supplementary Materials

Accompanying the manuscript:

**Computing the Expected Value of Sample Information Efficiently: Expertise and Skills Required for Four Model-Based Methods**

**Table of Contents**

# 1 PSA Matrix

We provide an example of a PSA dataset in Table S1 and an example of an augmented dataset in Table S2.

Table S1. An example of a PSA dataset from a traditional cost-effectiveness analysis

| Simulation | Input parameters $\theta=(\theta_1,...\theta_p)$ | | | | | | | | | | | | | Model outcomes | | | | | |
|---|---|---|---|---|---|---|---|---|---|---|---|---|---|---|---|---|---|---|---|
| sim | p.dr.t1 | hr.dr.t2 | ... | hr.dr.T | p.tox.t1 | p.tox.t2 | ... | p.tox.T | u.ndr | u.dr | u.d.tox | c.dr | c.tox | ... $\theta_p$ | qaly.t1 | qaly.t2 | ... qaly.T | cost.t1 | cost.t2 | ... cost.T |
| 1 | 0.259 | 0.611 | | 0.580 | 0.346 | 0.241 | | 0.057 | 0.846 | 0.721 | -0.031 | 88155.000 | 23426.000 | | 8.596 | 8.815 | 9.046 | 37407.287 | 38421.603 | 31748.150 |
| 2 | 0.268 | 0.557 | | 0.553 | 0.392 | 0.202 | | 0.063 | 0.766 | 0.718 | -0.032 | 127608.000 | 27283.000 | | 7.893 | 8.146 | 8.317 | 37340.764 | 38763.773 | 31856.518 |
| 3 | 0.279 | 0.530 | | 0.533 | 0.367 | 0.210 | | 0.059 | 0.789 | 0.691 | -0.030 | 70841.000 | 21433.000 | | 8.063 | 8.362 | 8.574 | 37755.966 | 39365.557 | 31297.161 |
| 4 | 0.285 | 0.625 | | 0.559 | 0.406 | 0.224 | | 0.058 | 0.897 | 0.683 | -0.024 | 69346.000 | 36290.000 | | 9.164 | 9.352 | 9.637 | 37824.883 | 38591.263 | 32562.561 |
| 5 | 0.272 | 0.568 | | 0.588 | 0.421 | 0.223 | | 0.056 | 0.912 | 0.681 | -0.014 | 127813.000 | 41895.000 | | 9.304 | 9.490 | 9.764 | 38169.000 | 39025.428 | 32089.763 |
| 6 | 0.295 | 0.569 | | 0.538 | 0.358 | 0.186 | | 0.058 | 0.767 | 0.669 | -0.022 | 80210.000 | 37017.000 | | 7.830 | 8.098 | 8.106 | 37015.991 | 38314.785 | 32258.373 |
| 7 | 0.270 | 0.510 | | 0.615 | 0.365 | 0.239 | | 0.056 | 0.875 | 0.678 | -0.026 | 112770.000 | 28003.000 | | 8.858 | 9.123 | 9.377 | 37661.349 | 38690.152 | 31649.673 |
| 8 | 0.294 | 0.623 | | 0.532 | 0.376 | 0.201 | | 0.060 | 0.692 | 0.672 | -0.032 | 67454.000 | 48947.000 | | 7.270 | 7.442 | 7.631 | 38617.847 | 39671.819 | 33022.915 |
| 9 | 0.238 | 0.638 | | 0.532 | 0.364 | 0.266 | | 0.062 | 0.880 | 0.709 | -0.029 | 60469.000 | 43479.000 | | 8.908 | 9.196 | 9.445 | 37911.386 | 39127.860 | 34232.103 |
| 10 | 0.240 | 0.646 | | 0.563 | 0.406 | 0.236 | | 0.057 | 0.904 | 0.714 | -0.016 | 105465.000 | 28510.000 | | 9.199 | 9.453 | 9.658 | 37604.479 | 38785.650 | 33186.151 |
| 11 | 0.283 | 0.586 | | 0.516 | 0.383 | 0.184 | | 0.054 | 0.895 | 0.735 | -0.026 | 148110.000 | 41794.000 | | 9.036 | 9.403 | 9.650 | 38162.548 | 39909.570 | 32429.058 |
| 12 | 0.258 | 0.587 | | 0.558 | 0.420 | 0.214 | | 0.058 | 0.868 | 0.744 | -0.018 | 111303.000 | 21353.000 | | 8.894 | 9.076 | 9.270 | 38152.053 | 39013.707 | 32842.565 |
| 13 | 0.279 | 0.561 | | 0.543 | 0.416 | 0.233 | | 0.062 | 0.871 | 0.686 | -0.033 | 140005.000 | 25705.000 | | 9.010 | 9.134 | 9.393 | 37420.385 | 37995.135 | 32401.545 |
| 14 | 0.277 | 0.565 | | 0.574 | 0.361 | 0.213 | | 0.060 | 0.902 | 0.694 | -0.012 | 104599.000 | 47773.000 | | 9.144 | 9.432 | 9.680 | 37598.085 | 39008.810 | 32006.206 |
| 15 | 0.268 | 0.583 | | 0.564 | 0.383 | 0.263 | | 0.063 | 0.931 | 0.689 | -0.019 | 143150.000 | 33170.000 | | 9.417 | 9.672 | 9.860 | 38176.451 | 39434.117 | 34439.495 |
| 16 | 0.246 | 0.624 | | 0.583 | 0.374 | 0.229 | | 0.065 | 0.787 | 0.672 | -0.020 | 55317.000 | 34668.000 | | 8.063 | 8.304 | 8.471 | 37877.865 | 39144.156 | 32905.542 |
| 17 | 0.274 | 0.561 | | 0.577 | 0.429 | 0.202 | | 0.060 | 0.810 | 0.668 | -0.027 | 136863.000 | 35512.000 | | 8.280 | 8.573 | 8.727 | 37919.837 | 39625.260 | 32432.658 |
| 18 | 0.266 | 0.617 | | 0.557 | 0.382 | 0.200 | | 0.066 | 0.754 | 0.690 | -0.021 | 60116.000 | 20616.000 | | 7.694 | 7.995 | 8.205 | 37624.118 | 39293.243 | 31603.192 |
| 19 | 0.242 | 0.574 | | 0.561 | 0.376 | 0.202 | | 0.061 | 0.856 | 0.676 | -0.032 | 93546.000 | 31603.000 | | 8.578 | 8.957 | 9.193 | 37514.271 | 39384.790 | 31344.471 |
| 20 | 0.271 | 0.547 | | 0.574 | 0.393 | 0.247 | | 0.058 | 0.805 | 0.700 | -0.026 | 50136.000 | 43776.000 | | 8.242 | 8.494 | 8.795 | 37655.252 | 38972.651 | 32556.914 |
| ⋮ | | | | | | | | | | | | | | | | | | | | |
| S | 0.279 | 0.609 | | 0.528 | 0.379 | 0.187 | | 0.065 | 0.774 | 0.674 | -0.030 | 143124.000 | 29488.000 | | 7.922 | 8.203 | 8.426 | 37161.192 | 38835.434 | 32120.867 |

QALY, quality-adjusted life years; θ, set of inputs in a decision-analytic model; S, number of parameter sets that are drawn from p(θ) in PSA; T, number of considered interventions; θ=(φ,ψ), φ are model parameters for which we are aiming to collect further information and ψ are parameters that will not be directly informed by the proposed data collection; p.dr.t, input parameter probability of distant recurrence for strategy $t=1,…,T$; hr.dr.t, input parameter hazard ratio of distant recurrence for strategy $t=1,…,T$; p.tox.t, input parameter probability of treatment toxicity for strategy $t=1,…,T$; u.ndr, input parameter utility weigth for non distant recurrence health state; u.dr, input parameter utility weigth for distant recurrence health state; u.d.tox, input parameter utility weight decrement due to treatment toxicity; c.dr, input parameter cost of distant recurrence health state; c.tox, input parameter cost of treatment toxicity; qaly.t, model outcome QALYs for strategy $t=1,…,T$; cost.t, model outcome costs for strategy $t=1,…,T$.

Table S2. An example of an augmented PSA dataset

| Simulation | Net Monetary Benefit $NMB_t^{\theta s}$ | | | | | EVPPI | | | | |
|---|---|---|---|---|---|---|---|---|---|---|
| sim | nmb.t1 | nmb.t2 | nmb.t3 | ... | nmb.T | evppi.t1 | evppi.t2 | evppi.t3 | ... | evppi.T |
| 1 | 837379.2024 | 844093.4111 | 839956.8991 | | 872811.4284 | 17869394 | 17869395 | 17869396 | | 17869398 |
| 2 | 772156.026 | 777257.5094 | 774127.9884 | | 799852.4296 | 11904043 | 13520849 | 15341966 | | 17638804 |
| 3 | 794212.5081 | 798408.4714 | 795838.8635 | | 826061.4942 | 15825190 | 11897424 | 13640558 | | 17828821 |
| 4 | 887204.7727 | 897373.6603 | 891338.8589 | | 931131.8558 | 25542950 | 26722330 | 25498593 | | 25257633 |
| 5 | 902637.5524 | 910823.0158 | 905776.2968 | | 944293.661 | 24996041 | 27301229 | 25909340 | | 22441169 |
| 6 | 768072.3318 | 772776.5576 | 769836.6696 | | 778373.3856 | 15226170 | 16250735 | 15544873 | | 15939131 |
| 7 | 864748.9975 | 874621.0878 | 868637.0989 | | 906095.0606 | 27089461 | 26779586 | 24764875 | | 18314127 |
| 8 | 708984.7723 | 705628.3338 | 707717.4533 | | 730085.6291 | 14554043 | 13713229 | 12237636 | | 17502532 |
| 9 | 867043.0172 | 881687.0493 | 872713.9937 | | 910279.6656 | 24317725 | 23296525 | 20718406 | | 22275512 |
| 10 | 895840.5652 | 907690.7155 | 900492.9759 | | 932648.821 | 19922874 | 25692721 | 19283669 | | 25507408 |
| 11 | 899460.1624 | 902122.2035 | 900489.0025 | | 932585.6119 | 27335340 | 21925089 | 19021609 | | 19924847 |
| 12 | 868201.8478 | 869490.9773 | 868685.1929 | | 894150.6201 | 24746060 | 24521569 | 19951257 | | 18238040 |
| 13 | 864847.2778 | 875970.4598 | 869322.0842 | | 906931.8712 | 24289546 | 22274054 | 20124633 | | 23213854 |
| 14 | 904399.3702 | 905629.8556 | 904879.6573 | | 936043.5577 | 18754670 | 22322519 | 26505429 | | 26484267 |
| 15 | 912275.4667 | 929050.7264 | 918752.9752 | | 951513.1796 | 19337510 | 26245599 | 26164331 | | 27418776 |
| 16 | 787142.5029 | 792503.1899 | 789222.9017 | | 814214.0454 | 12775933 | 17029562 | 17557082 | | 12420810 |
| 17 | 816896.277 | 819413.2747 | 817919.572 | | 840310.9478 | 12977702 | 12211846 | 16403519 | | 12321185 |
| 18 | 761366.6911 | 761833.7553 | 761549.6891 | | 788945.9899 | 12273772 | 15049874 | 14523964 | | 16973338 |
| 19 | 854949.0707 | 858167.5316 | 856167.4658 | | 887943.4643 | 23996038 | 18734306 | 18861141 | | 23795211 |
| 20 | 805175.773 | 811744.4666 | 807750.2668 | | 846916.9499 | 15165927 | 15216389 | 16261859 | | 13213453 |
| ⋮ | | | | | | | | | | |
| S | 782384.6787 | 783123.99 | 782674.7414 | | 810480.3205 | 15147702 | 14028967 | 12660209 | | 13664229 |

θ, set of inputs in a decision-analytic model; S, number of parameter sets that are drawn from p(θ) in PSA; T, number of considered interventions; θ=(φ,ψ), φ are model parameters for which we are aiming to collect further information and ψ are parameters that will not be directly informed by the proposed data collection; nmb.t, net monetary benefit for strategy $t=1,…,T$ ; evppi.t, expected value of partial perfect information for strategy $t=1,…,T$.

## 2 Prior Effective Sample Size Algorithms

This section proposes three alternative methods for calculating the prior effective sample size. The effective sample size $n_0$ is the number of patients that would have to provide data to generate the "amount" of information in the prior.

### 2.1 Direct Estimation

In some settings, $n_0$ can be found from the parameters prior distribution. For example, $n_0$ for a beta prior $Beta(a,b)$ coupled with a binomial likelihood $x \sim Bin(n,p)$ is equal to $a+b$. This comes from conjugacy as the posterior distribution is $Beta(a+x, b+(n-x))$. This is the case for specific conjugate pairs listed in Table S3.

Table S3. Identifying $n_0$ from parameters' prior distribution

| Prior | Likelihood | Effective Sample Size |
|---|---|---|
| $Beta(a,b)$ | Binomial | $a+b$ |
| $Gamma(a,b)$ | Exponential | $a$ |
| $Gamma(a,b)$ | Poisson | $\dfrac{1}{b}$ |
| $Normal(a,b)$ | Mean in Normal with known variance $\sigma^2$ | $\dfrac{\sigma^2}{b}$ |
| $InverseGamma(a,b)$ | Variance in Normal with known mean | $a$ |

### 2.2 Calculating a Summary Statistic

This method uses a Gaussian approximation to compute $n_0$ based on a summary statistic $W(\mathbf{X}_s)$. Specifically, the data must be simulated and summarised following the method described in the *Specification of data generating distribution* section. We can assume that this summary statistic is approximately Gaussian and that the prior for the parameter of interest is also approximately Gaussian. In this setting,

$$Var\,(W(\mathbf{X}_s)) = \frac{\sigma^2}{n} + \frac{\sigma^2}{n_0},$$

where $\sigma^2$ is an unknown variance. However, we know that $Var\,(\phi) = \dfrac{Var\,(\mathbf{X})}{n_0}$ so we can solve for $n_0$;

$$\hat{n}_0 = n\left(\frac{Var\,(W(\mathbf{X}_s))}{Var\,(\phi)} - 1\right),$$

where $Var(W(\mathbf{X}_s))$ and $Var(\phi)$ are the variances of the summary statistic and the prior, respectively. Therefore, if we have calculated the summary statistic by summarising a dataset of size $n$, we can compute the $n_0$ using this formula.

However, in these cases, we need to compute a summary statistic that is "on the same scale" as the parameter. For example, the prior in a beta-binomial is typically a probability between zero and 1, while the binomial likelihood is generally the distribution of the number of successes. Therefore, the summary statistic must convert the data into the proportion of successes, which can be easily achieved by dividing the number of successes by the total sample. In some setting this may be non-trivial. For these settings, we propose an indirect Markov chain Monte Carlo (MCMC) approach.

## 2.3 Using MCMC

In this case, we calculate the posterior mean for the parameter of interest $\boldsymbol{\phi}$ for each simulated value of $\mathbf{X}_s$, denoted $\mu_{\boldsymbol{\phi}}^s$. The prior effective sample size can then be estimated from the *variance* of $\mu_{\boldsymbol{\phi}}^s$ using the following formula:

$$\hat{n}_0 = n\left(\frac{Var(\phi)}{Var(\mu_{\boldsymbol{\phi}})} - 1\right).$$

More information about calculating $n_0$ can be found in [1] alongside code to estimate $n_0$ within R.

## 3 Supplementary References

S1. Jalal H, Alarid-Escudero F. A Gaussian approximation approach for value of information analysis. Med Decis Making. 2018; 38: 174-88.